\documentclass[letterpaper]{article}
\usepackage{iccc}

\usepackage{times}
\usepackage{helvet}
\usepackage{courier}
\usepackage{times}
\usepackage{soul}
\usepackage{url}
\usepackage[hidelinks]{hyperref}
\usepackage[utf8]{inputenc}
\usepackage[small]{caption}
\usepackage{graphicx}
\usepackage{amsmath}
\usepackage{amsthm}
\usepackage{booktabs}
\usepackage{algorithm}
\usepackage{algorithmic}
\usepackage[switch]{lineno}

\title{Design Considerations for Automatic Musical Soundscapes of Visual Art for People with Blindness or Low Vision}

\author{Stephen James Krol \\
        SensiLab, Monash University \\
        Melbourne, Australia \\
        stephen.krol@monash.edu
        \And 
        Maria Teresa Llano \\ 
        SensiLab, Monash University \\
        Melbourne, Australia \\
        Teresa.Llano@monash.edu
        \And
        Matthew Butler \\
        Monash University \\
        Melbourne, Australia \\
        matthew.butler@monash.edu
        \And 
        Cagatay Goncu \\
        Monash University \\
        Melbourne Australia \\
        Cagatay.Goncu@monash.edu}

\begin{document} 
\maketitle
\begin{abstract}
Music has been identified as a promising medium to enhance the accessibility and experience of visual art for people who are blind or have low vision (BLV). However, composing music and designing soundscapes for visual art is a time-consuming, resource intensive process - limiting its scalability for large exhibitions. In this paper, we investigate the use of automated soundscapes to increase the accessibility of visual art. We built a prototype system and ran a qualitative study to evaluate the aesthetic experience provided by the automated soundscapes with 10 BLV participants. From the study, we identified a set of design considerations that reveal requirements from BLV people for the development of automated soundscape systems, setting new directions in which creative systems could enrich the aesthetic experience conveyed by these. 
\end{abstract}

\section{Introduction}

People who are blind or have low vision (BLV) have a desire to visit museums to appreciate visual art \cite{HandaKozue2010Iopn} and they want to do so independently, without always relying on friends and family \cite{AsakawaSaki2018TPaF,Holloway}. Audio guides and tours are frequently employed as tools to enhance the accessibility of visual art, but often focus solely on description, which can restrict the aesthetic enjoyment for BLV users \cite{NGV,li2023understanding,AsakawaSaki2019AIaI,Holloway}. Tactile methods are another popular choice for interaction amongst BLV individuals as they allow users to independently explore various low level features of the artwork; however, the level of descriptive information is still limited \cite{multimodal}. This has driven the development of multimodal systems to provide a more complete experience for BLV users \cite{multimodal,multimodal2}. The outcomes from these systems are encouraging and pave the way for new modalities that can contribute to enhancing these experiences.

One aspect of visual art accessibility that could be improved upon is the access to the aesthetic experience \cite{Martins,RectorKyle2017EAEP}. We define an aesthetic experience in visual art as: 
\begin{quote}
\textit{A subjective experience that involves the way people appreciate different aspects of an artwork; for instance, beauty, colour, style, etc. Typically an aesthetic experience will invoke an emotional response that resembles a connection that you feel with the artwork, even by placing our own experiences in it.}
\end{quote} 
While both audio descriptions and tactile plates can provide some level of aesthetic experience \cite{li2023understanding}, their descriptive nature is limiting.

One approach that has shown promise is the use of music and soundscapes to convey mood and setting \cite{RectorKyle2017EAEP,Holloway}.
However, the process of making musical soundscapes can be time-consuming and resource intensive leaving a gap for automatic systems that can make adoption of this technique more viable. While some systems like this already exist \cite{Fink,Scene2Wav}, none have been thoroughly evaluated by BLV people, resulting in little understanding of what is required from an automatic musical soundscape system to ensure positive experiences for BLV individuals.

In this paper, we set out to build a foundation for AI-based musical soundscape systems for BLV users by understanding (1) the effect that these soundscapes have on the experience of BLV users and (2) important design features, such as building a narrative and using historically accurate sound effects, that should be included to ensure appropriate experiences. 
To achieve this, we built a prototype system, informed by previous BLV work \cite{RectorKyle2017EAEP}, that generates musical soundscapes of visual art and ran a qualitative study with 10 blind/low vision participants. 

The main contributions of this paper are as follows:
\begin{enumerate}
    \item An in-depth qualitative study, with BLV participants, exploring the impact of automatic musical soundscapes of visual art on their overall aesthetic experience.
    \item Design considerations, with a focus on the aesthetic experience, for automatic musical soundscape systems.
\end{enumerate}

\section{Background}

\subsection{Accessibility of Visual Art}
\subsubsection{Current Accessibility Methods}
The predominant method used to improve the accessibility of visual art is audio descriptions provided by either tour guides or pre-recorded devices \cite{AsakawaSaki2019AIaI,Holloway,li2023understanding}. These descriptions allow users to access a wide-array of information pertaining to the artwork from low-level features to general themes. However, while these technologies improve the accessibility of the artwork, their descriptive nature can limit access to the aesthetic experience for different BLV groups \cite{li2023understanding}.

Tactile methods are another popular method of interaction for BLV people \cite{Holloway,stangl2015transcribing,bearman20113d} and have been shown to be more effective at eliciting aesthetic pleasure in individuals born blind \cite{li2023understanding}. However, 
they are limited by their inability to provide iconographic and oconological information \cite{multimodal}. This has lead to the development of multimodal systems that combine audio guides and tactile models to allow for wider access to visual art \cite{multimodal,multimodal2}. These systems have demonstrated promising results and highlight the importance of investigating multimodal experiences to enhance aesthetic appreciation among BLV users \cite{Holloway}. 

\subsubsection{Accessibility of the Aesthetic Experience}
A study by Martins (\citeyear{Martins}) demonstrated the importance of making the aesthetic experience accessible to BLV patrons by running interactive workshops designed to take art and convey an aesthetic experience to blind participants utilising techniques such as collage and concealment. However, a limitation of the workshops is that they require a lot of resources to run, affecting the ability of museums to continuously operate them throughout the day, and highlighting the need for on-demand methods. 
Rector et al. (\citeyear{RectorKyle2017EAEP}) created an on-demand system named Eyes-Free Art. The system utilised a proximity sensor to detect the user and provided different levels of interaction depending on how close they were to the painting. The different levels either played background music, a sonification of the artwork, sound effects relating to the work or audio descriptions, all were designed to convey an aesthetic experience to the user. 
While the work by Rector et al. (\citeyear{RectorKyle2017EAEP}) has the capacity to introduce new dimensions of experience for BLV individuals, the resource-intensive nature of crafting such experiences limits its potential as a tool for accessibility, highlighting the need for automatic approaches.

\subsection{Generating Sound from Images}
Traditionally, research that aims to generate sound from images involves directly mapping visual features to audio \cite{CavacoSofia2013CSft,Sonicphoto,Paint2Sound,Photosounder}. While this can help BLV patrons better understand certain features \cite{CavacoSofia2013CSft}, these techniques provide little descriptive or aesthetic features to listeners.

To create a different experience, Zhao et al. (\citeyear{SichengZhao2014Ebim}) utilised the principles-of-art \cite{Collingwood} to musicalise images using emotion. They derived an emotive score that would be used to find an appropriate musical track in an emotionally annotated database \cite{schmidt2011modeling}. Emotion is an important part of the aesthetic experience and the authors succeeded in conveying it through music. However, they relied on an existing database that could potentially match two paintings to the same piece of music instead of creating unique experiences for each artwork. 

Sergio \& Lee (\citeyear{Scene2Wav}) also utilised emotion to musicalize images and built a deep learning model called Scene2Wav which generated emotional music that matched short movie scenes. However, as with similar work \cite{tan2020automated}, it could only convey the emotion of the scene and did not provide any contextual information, an important element for BLV patrons \cite{RectorKyle2017EAEP}. Fink et al. (\citeyear{Fink}) provided contextual information by sonifying both low-level features, such as colour, and high-level features, such as the scene of the image.  However, no BLV patrons evaluated the system and in the samples provided by the authors, sound effects were ambient background noises and did not include information on foreground objects.

\section{Prototype System Description}
\begin{figure}[h]
    \centering
    \includegraphics[width=0.5\textwidth]{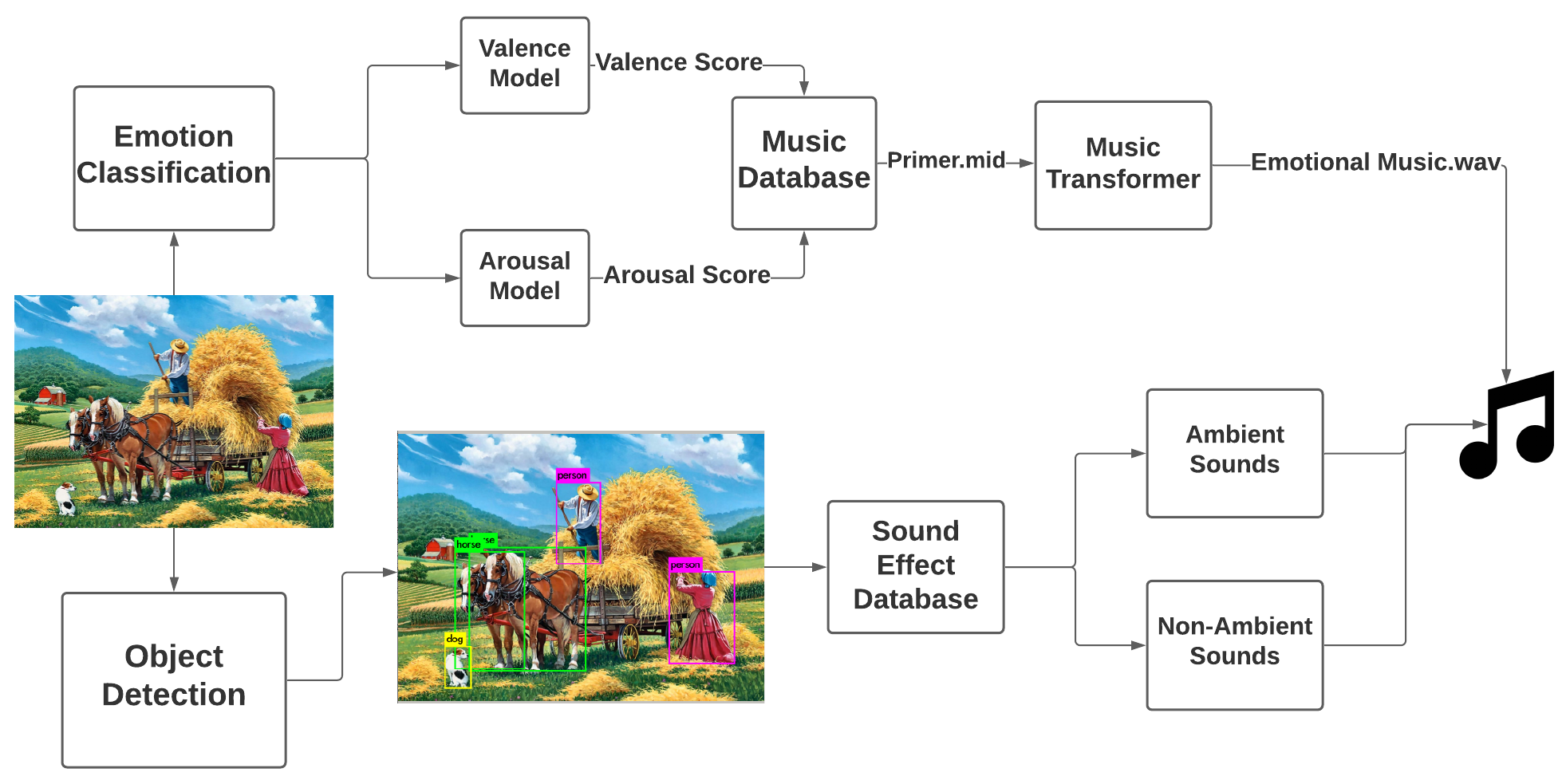}
    \caption{System Diagram: Depicts the various tasks performed by the system to generate the output music.}
    \label{fig:system_diagram}
\end{figure}

Like the work above, our musical soundscapes attempt to create an aesthetic experience for the listener by conveying the perceived emotion of the painting through the generated music. Additionally, our prototype system conveys contextual information through both foreground and background sound effects. This was done to mimic some of the features present in the successful human curated Eyes Free Art experience \cite{RectorKyle2017EAEP} designed for BLV users. To the best of our knowledge, there is no other system that does all of this automatically. It is worth noting that this system is designed for representational art, work relating to abstract art can be found here \cite{framingthroughmusic}.
The current version of our prototype system generates musical soundscapes from art in two stages: music generation and sound effect curation. A diagram of our system can be seen in figure \ref{fig:system_diagram}.

\subsection{Stage 1: Generating Music with Emotion}
Emotion is an important part of the aesthetic experience and it is important that the system can generate emotive music that represents the painting. Building off work presented here \cite{framingthroughmusic}, our system generates emotive music by (1) classifying the perceived emotion of the painting in the valence-arousal scale, (2) finding a musical primer that matches this emotion and (3) conditionally generating new music with this primer. The following subsections describe this in more detail. 

\subsubsection{Emotion Classification} 
\begin{figure}
    \centering
    \includegraphics[scale=0.22]{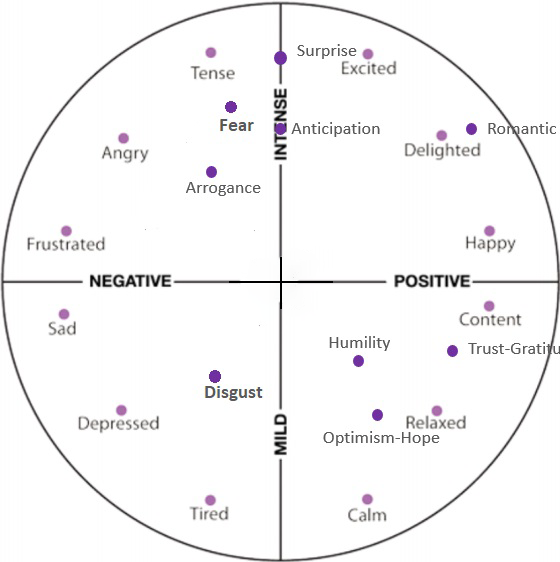}
    \caption{Emotional Mappings: The above figure shows the mappings used to convert the categorical emotions in WikiArt to emotions in the Valence-Arousal scale.}
    \label{fig:emotion_mappings}
\end{figure}

To classify emotion in the painting we utilise transfer learning from the Imagenet InceptionV3 \cite{szegedy2015rethinking} model training on the WikiArt emotions dataset \cite{LREC18-ArtEmo}. WikiArt emotions contains 4000 pieces of art that have been emotionally annotated by various observers with a total of 20 emotions. Some emotions in the dataset only applied to less than 1\% of the paintings, therefore to simplify training these emotions and their corresponding paintings were removed. Furthermore, the emotions labelled as `trust' and `gratitude' were combined as they have similar valence and arousal scores. The final set of emotion labels used in this study were happiness, surprise, anticipation, trust-gratitude, humility, fear, sadness, optimism, love, disgust, arrogance and anger. An important step in preparing the dataset for training was mapping the categorical emotions to their respective values in the Valence-Arousal model \cite{russell1980circumplex}. This is because the music database utilised by the system annotated emotions in the valance-arousal scale. Mapping the emotions involved analysing psychology literature and determining a consensus for each emotion \cite{JinXuecheng2005AESM,de-bruyne-etal-2020-emotional,HussainM,WangLing2021MERB,SellersMichael2013Tact}. We acknowledge that this is still an area of active research and there are arguments for and against the mappings we present in figure \ref{fig:emotion_mappings} (original figure from \cite{ferreira_ismir_2019}); however, we believe that each emotion is generally categorised appropriately.

To predict emotion, we decided to split our model into two separate networks: one to predict valence and another to predict arousal. This was because the WikiArt dataset is biased towards positive valence and high arousal emotions, and using a single model made it difficult to balance classes during training. Splitting the model into two networks rectified this problem, allowing us to effectively use undersampling to ensure balanced training. The training and testing losses for the valence model are 0.09 and 0.12 respectively and for the aoursal model 0.1 and 0.15 respectively.

\subsubsection{Emotionally Annotated Music Database}
All tracks used as a primer for the music transformer were from the VGMIDI annotated database \cite{ferreira_ismir_2019}. The primer is a short piece of symbolic music that the music transformer takes as input and uses to start the music generating process. 

\subsubsection{Music Generation} 
Once an appropriate track was found, Magenta’s music transformer was used to generate the music. Music transformer is an attention-based neural network that attempts to generate music with coherent long-term structure. We utilised a ported script from the Google Music Transformer notebook \cite{piano_transformer} with the melody\_conditional\_model\_16 weights, to automatically generate midi files using the Music Transformer. 

\subsection{Stage 2: Incorporating Contextual Information}
In addition to emotive music, the system also provides information relating to the context of the art, such as the setting, that should also be communicated to the user. Stage 2 does this by (1) Detecting objects within the painting, (2) Retrieving relevant sound effects from a database and (3) combining the sound effects and music.

\subsubsection{Object Detection}
The Yolov3 algorithm \cite{redmon2018yolov3} was used to detect objects within the painting. YoloV3 can only detect 80 different classes, which limits the system's ability to convey contextual information to the user. However, whilst other models, like YoloV9000 \cite{redmon2016yolo9000}, can detect more classes, these models would often produce incorrect results. Therefore, the system sacrifices quantity over quality. It is also worth noting that these algorithms are designed to detect real-world objects and while it is still able to detect painted objects, we would expect this to fail on more abstract representations.

\subsubsection{Sound Effects Database}
The final step in the process involves finding sound effects that match the objects detected in the painting. The system utilises a custom-built database that contains sound effects for 51 different objects. Sound effects are classified as either ambient or non-ambient. Ambient sound effects are intended to play throughout the entirety of the song and are used to convey the setting of the painting. An example of an ambient sound effect is the sound of rain. A non-ambient sound effect is meant to describe the foreground objects within the painting. An example of a non-ambient sound effect would be the sound of a cow mooing.

\subsection{Pilot Study}
\begin{figure}[t]
    \centering
    \includegraphics[scale=0.15]{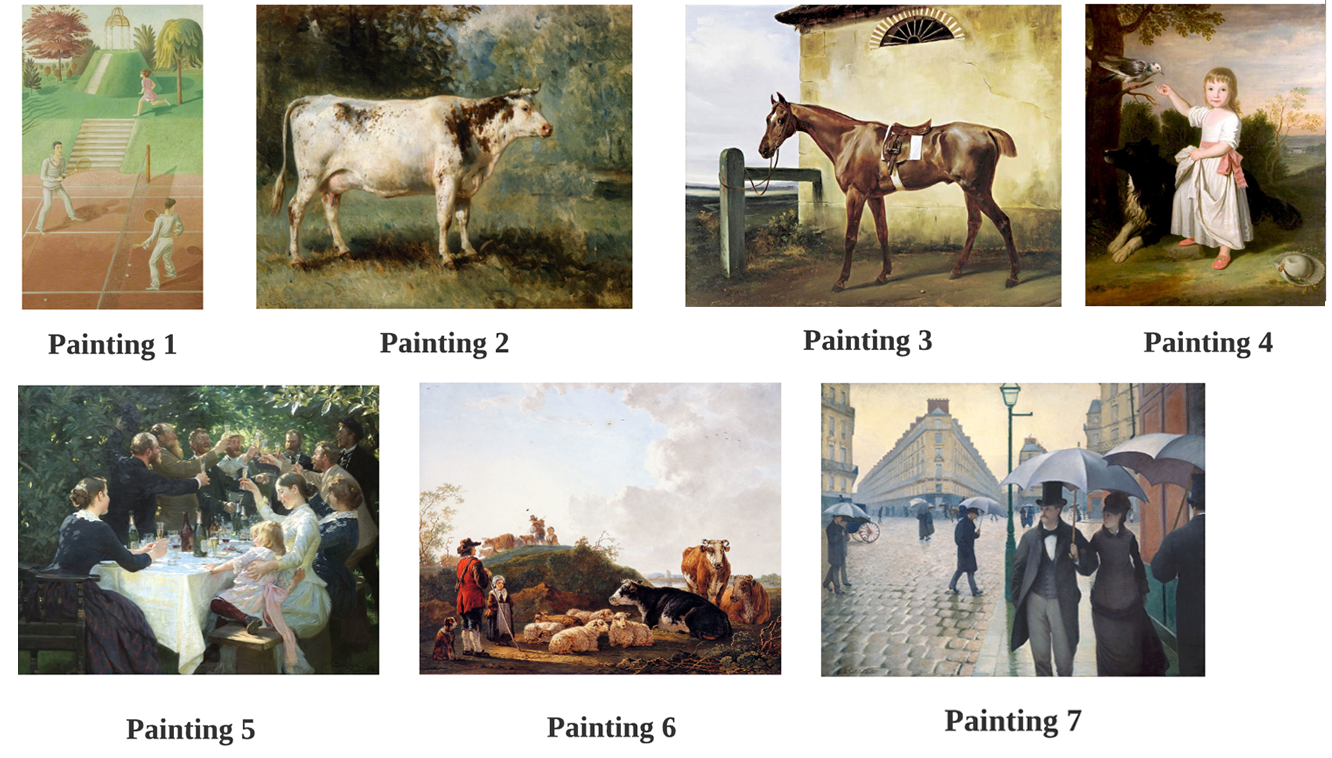}
    \caption{Paintings used in the study. 1: Tennis Triptych Centre Panel, 2: A Cow in a Landscape, 3: A Saddled Horse Tied to a Fence, 4: James Abercrombe of Tullibody esq, 5: Hip Hip Hurrah, 6: Herdsmen with Resting Cattle, and 7: Paris a Rainy Day.}
    \label{fig:examples}
\end{figure}

We first conducted a preliminary study with sighted and low vision participants in order to validate the suitability of the system with people of different levels of visual acuity. 
The study involved participants experiencing soundscapes generated by the system for seven different paintings (see Figure \ref{fig:examples}). These paintings depicted a variety of scenes with diverse objects/subjects, conveyed different values of valence and were not part of the system's training set. 

The study was conducted amongst 28 sighted and 5 low vision participants (n=33) -- recruitment of BLV participants proved difficult as the pilot study was carried out during Covid-19 lockdown -- 22 participants identified as male, 10 as female, and 1 as non-binary. Participants' age varied from 18 to 55+. For the pilot study we conducted a survey that consisted of 4 questions that aimed to gauge the participants' emotional perception (from a list of options), level of enjoyment when interacting with each painting (7 Likert scale), how well the system incorporated foreground information in its output (open ended question), and the degree of accuracy with which the generated music represented the artwork (7 Likert scale). 
The results were mostly positive with participants stating that their experience was pleasant 70\% of the time with a median score of 5 and that the music was representative of the artwork 75\% of the time with a median score of 6. The emotion perceived by participants during their experience matched the annotated emotion of the painting 60\% of the time, and participants correctly identified sound effects relating to the painting 92\% of the time, with all low vision participants correctly identifying the objects within the musical soundscapes. Additionally, BLV participants found the soundscapes more pleasant with a median score of 6 compared to 5 for non-BLV participants. This indicated that the output of the system could be used reliably to test the potential of automatic musical soundscape systems for BLV individuals. 

\section{In-depth Qualitative Study}
The in-depth qualitative study was conducted with BLV participants (n=10) in order to identify relevant design considerations for using automatically generated musical soundscapes as an accessibility method for visual artworks. 

\subsection{Participants} 
\begin{table}[t]
    \centering
    \setlength{\tabcolsep}{3pt}
    \begin{tabular}{c|cp{3cm}c}
    \hline
        {\bf Participant} & {\bf Age} & {\bf Visual Acuity} & {\bf MSP} \\
    \hline
        L1 (F) & 25-34 & Snow tunnel vision, blind spots and night blindness & 78 (raw: 98) \\ 
        L2 (F) & 35-44 & Can see light, dark and colours & 93 (raw: 117) \\ 
        L3 (M) & 55-64 & Partial sight & 70 (raw: 88) \\ 
        L4 (F) & 45-54 & Central vision on the left, no vision on the right & 65 (raw: 82) \\ 
        L5 (M) & 45-54 & Light perception in one eye, no shapes, no colour & 79 (raw: 99) \\ 
        B1 (F) & 35-44 & 2\% vision, light and dark perception & 61 (raw: 77) \\ 
        B2 (F) & 35-44 & Blind since birth & 77 (raw: 97) \\ 
        B3 (F) & 18-24 & Blind since birth & 63 (raw: 79) \\ 
        B4 (F) & 35-44 & Blind since birth & 84 (raw: 106) \\ 
        B5 (M) & 45-54 & Blind since birth & Unknown \\ 
    \hline
    \end{tabular}
    \caption{In-depth qualitative study: Participants' demographic information (F=Female, M=Male). The Music Sophistication Percentile (MSP) was obtained using the Goldsmiths Musical Sophistication Index (\url{https://shiny.gold-msi.org/gmsi_toplevel/.}). The table shows the percentile for the raw scores of the General Music Sophistication factor; 
    e.g. L1's percentile of 72 means that the L1's self-assessed level of music sophistication is higher than 72\% of the population.}
    \label{tab:participantsDemographics}
\end{table}

From the ten participants, five of them were blind and five had low vision (see Table~\ref{tab:participantsDemographics}). From the blind participants, 4 were blind from birth and 1 lost their vision in their adulthood. From the low vision participants, diverse levels of vision were described as specified in Table~\ref{tab:participantsDemographics}. 

As music is at the heart of the approach, we wanted to understand the level of music sophistication of the participants for our analysis. To obtain a more objective measure of this, we used The Goldsmiths Musical Sophistication Index (Gold-MSI) \cite{Mllensiefen:2014}, in particular we focused on the General Sophistication factor as the aim was to have a high level understanding of the participants' musical competence. The results of the self-assessment show different levels of music sophistication in our participants, with a mean of 74.4 and median of 77. In addition, 3 participants considered themselves musicians and 8 participants have engaged in regular, daily practice of a musical instrument for more than 10 years. These results demonstrate a high level of musical competence from our pool of participants, giving higher confidence to the results of the study. 

\subsection{Procedure}
The study consisted of an online semi-structured interview with each of the participants over zoom. Participants were compensated with the equivalent to US\$64 for their time and feedback. The interviews lasted around 1 hour each and were divided into three parts:
\footnotesize 5
\begin{enumerate}
    \item \textbf{Understanding current interactions with visual artworks}: Participants were asked questions about their current experience with visual art and were also introduced to the definition of Aesthetic Experience provided earlier in the paper.
    \item \textbf{Experiencing Visual Artworks through Music} : Participants experienced the musical soundscapes alongside the art shown in figure \ref{fig:examples} via shared screen through zoom\footnote{\url{https://www.youtube.com/playlist?list=PLJXhSHZOX4QwkaqoS-o-cTdc90dM_aFKh}}. For each artwork we either provided a short description first and then showed and played the piece, or vice-versa. This was done randomly in order to reduce biases. The descriptions used in the study were simple and factual in nature, the purpose of this was to limit the amount of information given as we were evaluating the effectiveness of the sound effects and emotional content of the musical pieces, as well as not to bias participants with our own subjective interpretations.
    \item \textbf{Contrast with existing accessibility methods}: Participants were asked questions about the musical soundscapes and how they compare to existing methods such as audio guides and tactile plates.
\end{enumerate}
 
\subsection{Thematic Analysis}

Transcriptions from the semi-structured interviews were coded using Braun and Clarke thematic analysis methods \cite{BraunVirginia2006Utai} to identify common themes. Two researchers independently coded each participant's interview transcripts to produce a list of preliminary codes, before comparing and combining their analyses and settling on the final set of themes.

\section{Results} 

All participants reported to have visited art galleries and museums at some point in their lives; however, as found in previous studies \cite{li2023understanding,Holloway}, {\em gallery inaccessibility}, which manifests in various forms, was deemed as the major challenge for experiencing the artwork and achieving an aesthetic experience. For our participants, the most common method used to access art and an aesthetic experience was through auditory descriptions of the painting. However, participants repeatedly stated that tactile methods were more enjoyable and brought them closer to {\em `their own aesthetic experience'}. From these methods, participants' main critique was on the amount and type of information conveyed, which was often limited, focusing on certain aspects of the artwork while leaving others out (such as the painting's style, the artist's background). 

\subsection{Themes arising from thematic analysis}

\begin{figure}    
    \centering
    \includegraphics[scale=0.12]{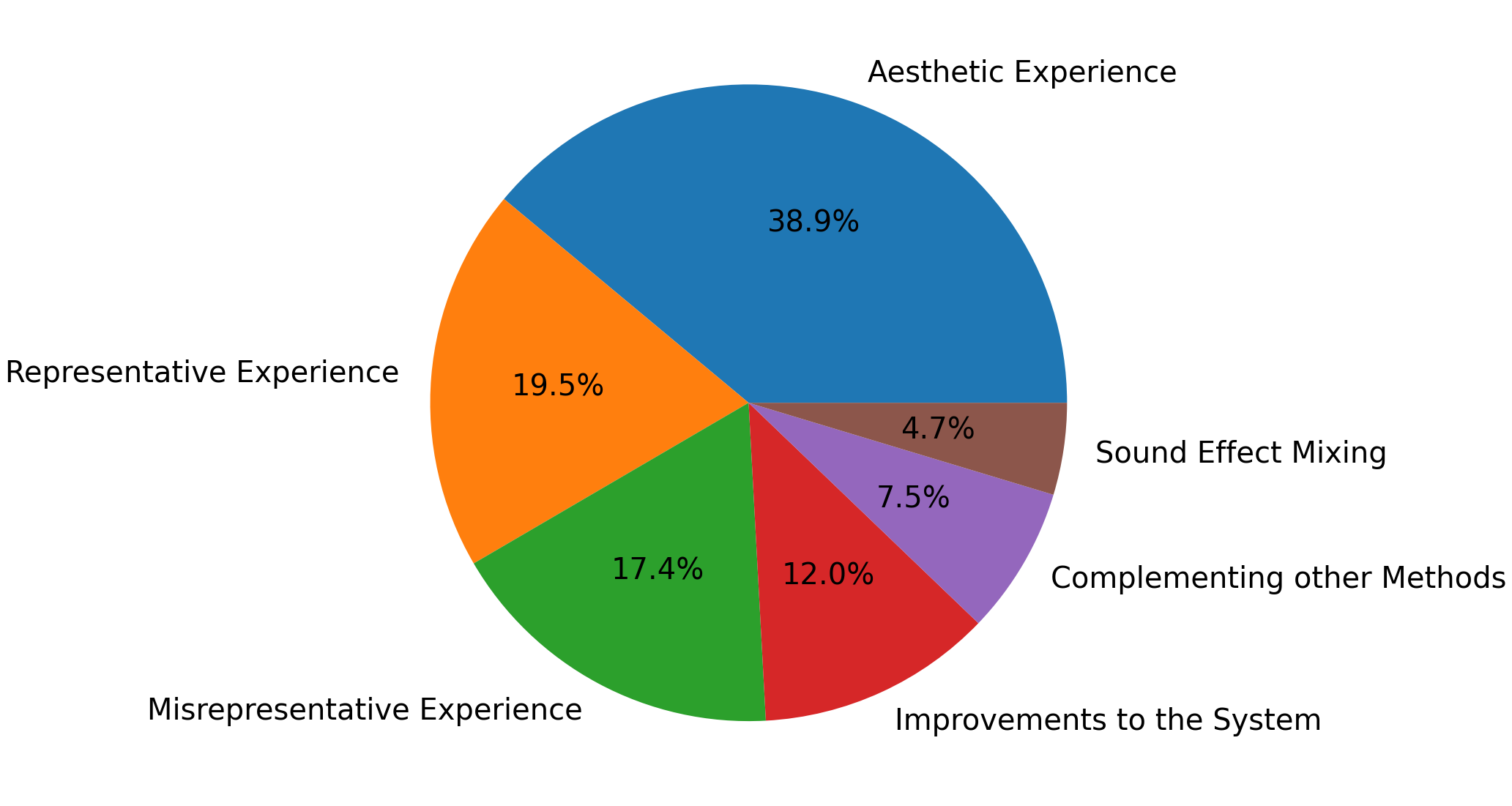}
    \caption{Themes identified through the thematic analysis of participants' interviews.}
    \label{fig:sunburst}
\end{figure}

The set of main themes identified from the thematic analysis are shown in Figure~\ref{fig:sunburst}. In the remainder of the section we will describe these themes and the codes within them. For the purpose of our analysis we will refer to {\em musical soundscapes} when we talk about the entire audio (i.e. music plus sound effects), {\em music} when we talk about the musical component of the experience, and {\em sound effects}, when we talk about the sounds that represent objects and subjects in the artworks. 

\subsubsection{Theme. Aesthetic experience:} 
The focus of this theme is on how the musical soundscapes affected the participants' aesthetic experience. The findings indicate that these had a predominantly positive effect, with 70\% of references being positive and only 8\% being negative.

When describing their positive experiences, participants expressed how the soundscapes {\em helped build a connection} with the artworks through the music by setting the tone and mood of the piece ``there's something about music, that it sets the mood ... very, very quickly, within those few seconds'' (B1), and through the sound effects by helping them create focal points. For instance, when describing their experience with {\em James Abercrombe's} piece (Painting 6), L3 mentioned how the sound effect of the dog barking got their attention as if saying ``I'm lower to the ground. But I'm still here''. This was echoed by other participants and was described as providing ``a more rounded perspective'' (L3) as it not only brought their attention to elements of the artworks that may be less perceptible or predominant, but did so in a way that establish some sort of communication from the artwork to the BLV patron ``the dog might be saying, hey, don't forget me'' (L3).

Participants also expressed enjoyment on the effect of the musical soundscape {\em telling a story}, which resulted in more holistic experiences, ``a journey that you kind of, you kind of go on'' (B1), or by evoking memories or stories; e.g. ``I love the ones that tell a story. From an experience point of view. So probably like that Paris scene or the Happy Birthday scene. Even though that would not necessarily be my favourite painting, that would be my favourite experience because then there's something happening, there's a story'' (B1). Participants associated this sentiment with the fact that the soundscapes conveyed a dynamic aspect to the artworks, through movement and activity, that created the atmosphere of a sequence of events, and that evoked different emotions in the listener: ``you use the music and the sound effects to fuel that image, which then makes the art piece either more fun, more terrifying, more sad, depending on what the piece is trying to do'' (L5).

In contrast, exploration of the soundscapes' negative effects observes participants noting that the music was sometimes distracting and superfluous, it didn't match the sound effects, or the choice of instrument was wrong for the piece. A common factor here, was that participants expected the music to convey more in-depth information, such as the cultural background of the place where the artwork took place; e.g. ``maybe a slide guitar or a bottleneck slide, because then you get a bit of the western feel'' (B4). 

Analysis of this theme also revealed {\em external factors} that play a crucial role in shaping how people perceive and respond to the soundscapes. This included {\em diversity of visual acuity}, referring to the extent of participants' vision and the onset of any vision loss and meant that participants had different points of reference to ground the musical soundscapes; for instance, L4 noted, when referring to the 'Tennis Triptych Centre Panel (painting 1)', ``I had full vision until I was 25 ... because I know what things look like ... I know what Wimbledon looks like ...  that music initially is going to take me to something I've previously seen''. Another of the factors identified in this theme was {\em subjective taste} and demonstrated how subjectivity can affect the experience of a particular musical soundscape. For instance, one of the participants showed a positive inclination towards pieces that included animals sound effects while another one was moved by sound effects that evoked the image of being indoors during a rainy day.

\subsubsection{Theme. Misrepresentative Experience:} 
This theme delves into instances where the participants' perception of the musical soundscape differed from the visual representation of the artwork. This disparity mostly arose in cases where participants deemed there to be {\em inaccuracies in the sound effects}, while in other cases it was attributed to perceived {\em inaccuracies in the music}. 

In the first case, participants pointed out occasions in which the sound effects portrayed completely different objects/subjects; e.g. ``I guess the rain kind of sounded a bit like static'' (L1), the sound effects portrayed objects/subjects with features distinct from those in the artwork; e.g. ``no, definitely not a pigeon ... something small, a lot, a lot, a lot smaller'' (B4), or the sound effects lacked information on the action(s) that the objects/subjects were involved in; e.g. ``they [tennis balls] don't stay in the air that long, they fall out of the air pretty quick'' (L5).

In the second case, where the inaccuracy was due to the music, participants highlighted that the mood set by the music did not match the mood set by the artwork (and artwork description), particularly when reflecting about the objects/subjects in it and/or their actions;  e.g. ``It's not quite what I expected, especially if it's a racehorse ... it's a bit slow and a little bit, not sombre, but a bit serious'' (B1).
Additionally, participants reported occasions when the music failed to portray characteristics of the culture, or the period, being conveyed by the artwork; e.g. ``I think the music is extremely anachronistic'' (L2).
In some cases, slight misrepresentation was not a negative for participants, for example, two of the blind participants, B4 and B5 (born blind), suggested a level of acceptance as the medium would provide a way to have an aesthetic experience; for instance, B5 said ``I don't see it as a bad thing. I guess you know, it's my interpretation ... Yeah, I don't see it as a bad thing''. 

Despite this, misrepresentations still have a {\em negative effect} and should be avoided. In particular, in some of these misrepresentative instances, participants were able to identify correctly the object/subject being portrayed but noted that the sound effect wasn't a faithful representation of the actions being ``performed'' in the artwork, e.g. ``tennis balls when they hit the racket they do have a very distinct sound that I don't think any other kind of ball sport has'' (L1). Participants also emphasized the importance of accurately representing outdoor and indoor settings ``they sound as though they're in a room ... that doesn't sort of say outside to me'' (L5). 

Most of the participants also agreed that having a description before hearing the musical soundscape would help alleviate the negative impact caused by any inaccuracies in the sound effects; e.g. ``if I had the audio description first, I probably wouldn't even notice the cough because I would already have in my head `Oh, it's a it's a 12 year old girl, or an eight year old girl''' (B1). 

\subsubsection{Theme. Representative Experience:} 
This theme focuses on instances in which participants' perception of the soundscape aligned with the visual representation of the accompanying artwork. 
Regarding the {\em accuracy of the music}, participants mainly pointed out when the music interacted well with the scene depicted in the artwork (or described by the researchers); e.g. ``Yeah, it's the kind of music that you would expect'' (L3), and provided an emotional anchor to it; e.g. ``that's cool, you know, it makes me happy to hear the artwork then seeing the image come up to me ...  it made it more easier to get to an emotional attachment to it and liking of it'' (L4). Participants even mentioned occasions in which the music was able to convey a closer representation of the scene in terms of the period and place where it was supposed to take place; e.g. ``Kind of like folk music, which you would expect sort of something that's a bit more, yes, something that's Dutch. That's really lovely'' (B1). Finally, participants highlighted the ability of the music to match the perceived mood; e.g. ``This is peaceful. This is probably a landscape'' (B1), ``to me that it sounded bright. I felt like there was some sunshine or something there'' (B5).

In addition, the {\em accuracy of the sound effects} was representative of occasions when the participants provided a very straightforward description of the objects/subjects depicted in the artworks' scenes; for instance, ``I'm assuming it's a tennis ball being hit'' (L1), ``people sound like they're having fun and the drinks are flowing'' (L2), ``I love the the swish of the umbrella going up and down'' (B4). A very interesting point made by the participants was the ability of the sound effects to create a narrative, or sequence of events, of the scene; e.g. ``hit of the ball, and then the ball landing and stuff ... it's certainly conjure up an idea of a game of tennis'' (L5). This was also observed in the code {\em accuracy of experience}, where participants were able to provide general faithful descriptions of the entire scene depicted in the painting, even when they were not provided with upfront descriptions of the artwork.

\subsubsection{Theme. Sound effect mixing:} 
This theme focuses on properties related to how the sounds in the musical piece were mixed. Broadly speaking, participants reported on different aspects related to the {\em introduction of sound effects}, and how they could be used to effectively convey information about the artworks and improve the aesthetic experience. In particular, participants highlighted different features such as the volume of the sound effects being too loud; e.g. ``a horse winning can get pretty loud'' (L1), the timing of the sound effects; e.g. ``when we're listening, and we're getting all that information in, we need to process it, if you give us too much information at once it answers all the questions at once'' (L4), and the number of sound effects needed; e.g. ``I would expect more sound effects the busier the painting is. If it is a simple painting, less sound effects'' (B1). A comment that persisted throughout the interviews, was that of {\em sound effects panning}, for which the participants noted that through stereo the musical piece could convey important information about location, space and actions.

\subsubsection{Theme. Complementing other accessible methods:} 
Participants suggested that relying solely on this medium as a form of aid to improve the aesthetic experience may not be sufficient in accommodating individuals with different levels of vision, and suggested the importance of this medium being used in conjunction with other accessibility methods. The theme particularly indicated that the {\em order} and {\em balance} of methods, would be both equally crucial for this. For instance, one blind participant stated ``if you've got nothing but a piece of music, I'm gonna make up my own piece of artwork ... I'm gonna keep guessing and guessing and guessing and therefore all I'm doing is guessing artwork'' (L4). However, participants also pointed out that our approach resulted in a better aesthetic experience than when they relied only on other methods; for instance, one of the participants stated that ``the description is really useful. You kind of go yep, there's a cow in that picture. But the music gives you a sense of, yeah, the colour, the lightness, the scenery. Is the cow having a good day, like you get that emotive experience. So I think, the description is good, but with the music, definitely a richer experience'' (B1). Balance was also associated with the amount of information that should be provided by each method; for instance, one of the participants stated that short descriptions would be enough ``like there is a horse joining here. He's facing forward. He's tied to a house. The house is the yellow brick casts. That's it sort of thing'' (B4), while other stated that ``audio descriptions can often give away the punch line ... it gives away the whole thing, and you don't get to be part of the mystery ... part of the enjoyment of art is identifying the work'' (L4). Regarding the {\em order of methods} in the interaction, some participants indicated a preference towards having the description first and the musical soundscape later, while others preferred the other way around.

\subsubsection{Theme. Improvements to the soundscapes:} 
This theme encompasses aspects of the approach, and the system, that participants suggested could be improved. Firstly, participants reiterated that the musical soundscape should convey more {\em contextual information} about the artwork, highlighting different ways in which the music and the sound effects could enrich the experience, particularly reflecting the style of the artwork, the period and culture where it takes place, and the actions that objects and subjects are involved with. Secondly, participants restated the importance of improving the {\em sound effects database} in order to minimise inaccuracies in the sound effects (such as the sound of people being indoors when they should be outdoors), and in order to avoid repetitive sounds across different images. 

\section{Discussion}
In general, automatic musical soundscapes added a more dynamic feel to the experience conveying the perceived emotion of the painting through the music and communicating information regarding the context and setting of the artwork through various ambient and non-ambient sound effects. Our study also provided various insights about the use of automatic musical soundscapes as a medium to make visual artworks more accessible. We discuss these next.

\subsection{Emotion and Subjectivity}
In general, participants described a matching emotion to that intended by the system's identified emotion from the artworks in the majority of cases. The results from the pilot study also supported this, with participants (sighted and low vision) reporting the experience matching the emotional annotation of the art 60\% of the time. This demonstrates the potential of the musical soundscapes to convey a faithful emotion. However, emotion is a very complex area of research, with numerous efforts being made to conceptualise different models of emotion \cite{posner2005circumplex,watson1988development,plutchik1991emotions}. In addition, research has demonstrated that the perceived emotion in a musical piece (i.e. the emotion intended by a piece) may differ to the induced emotion (i.e. the emotion actually felt by people) due to temporal or other circumstances \cite{gabrielsson2001emotion}. Being a subjective part of music, and art appreciation in general, it is natural that people may disagree with the emotions that the soundscapes were trying to convey, in the same way that people may disagree with emotions an artist is trying to convey with their work. Musical soundscapes are subject to a similar subjective evaluation as other methods of art appreciation; however, it is important for the soundscapes to try to convey a faithful emotion of the artworks they are representing to ensure BLV have access to an appropriate aesthetic experience. 

\subsection{Human-made vs Machine-based Musical Soundscapes}

Despite the limitations of the automatically generated musical soundscapes, all participants in the in-depth qualitative study said that they would use them if they were available in museums and art galleries. An automatic approach significantly reduces the costs required to generate soundscapes 
of visual artworks, which in turn can result in enhancing the aesthetic experiences for BLV patrons. For example, if a museum was to implement Martins workshops \cite{Martins}, they would need to pay the wages of the instructors and ensure that all personnel were trained. Additionally, approaches like Eyes-Free art \cite{RectorKyle2017EAEP} would require hiring someone to create the experience for each individual artwork. These costs could deter museums and art galleries from utilising these systems and restrict the number of artworks that are made more accessible. Furthermore, the human element also reduces the speed at which a musical piece can be created. Automatic systems could be used to improve these systems by substituting their limiting features. For example, rather than having human chosen music, or finding relevant sound effects for Eyes-Free art, an automatic approach could be used to generate them automatically. This not only reduces the time required to create a musical piece, but also makes it more feasible for museums or art galleries to incorporate an interactive system like Eyes-Free art into their collections, improving the accessibility of the aesthetic experience for BLV patrons.

\subsection{Design Considerations}
The following set of design considerations were identified from the study:

\subsubsection{Enhance comprehension through narrative elements}

An important observation from the participants during the interviews was that the soundscapes had the potential of enhancing their comprehension of the artworks when they were perceived as conveying a narrative or story -- even though this was not designed into the system. Additionally, in many cases this also had the effect of enhancing their aesthetic experience (as described in the theme {\em Aesthetic Experience}). Previous work \cite{li2023understanding} also identified {\em learning related stories} and {\em imagination} as the second and third main sources of enjoyment for BLV patrons when experiencing visual artworks. The non-prescriptive nature of musical soundscapes makes them a great medium to exploit these two sources of enjoyment, as participants felt that the musical pieces were able to add an extra dimension, partly by adding narrative elements that depicted a more complete picture of the visuals in the artwork. This may involve adding narrative elements to musical pieces, that even if not portrayed in the artworks, are intrinsically related to them. Some of the examples suggested by the participants included: the galloping of a racehorse (even if the horse is standing still), the sound of an opening gate (to signify that a person is in the picture) or the sound of wind on trees (to illustrate a field). 

\subsubsection{Provide controls to customize soundscapes:}
Overall, experiencing art is a unique and subjective experience that can vary greatly depending on individual preferences, cultural background, and personal experiences. As in previous work on the use of audio descriptions \cite{stangl2021going}, our in-depth qualitative analysis demonstrated the importance of individual preferences when using soundscapes as an accessibility method for visual artworks. 

With this in mind, specific to musical soundscapes, we identified the following factors that have the potential for customization:
\begin{itemize}
    \item length of the musical pieces,
    \item density of sound effects in a single musical soundscape
    \item order of interaction with other accessibility methods (such as audio descriptions), 
    \item use (or not) of repetitive sounds, and
    \item the amount of details in audio descriptions that are used in conjunction with soundscapes.
\end{itemize}

\subsubsection{Combine with other accessibility methods:}
As identified in previous work \cite{Holloway,li2023understanding}, a single accessibility method may fall short with regards to the complexities of representing visual artworks. This is also the case for musical soundscapes. Testing how this method could work with other accessibility methods was out of the scope of this work; however, participants offered insights about it, particularly in conjunction with audio descriptions as has been previously described (e.g. how much detail to add to audio descriptions, and whether to present them before or after the soundscapes) suggesting promising future avenue for this work.

\subsubsection{Employ historical and contextually diverse sound effects and music to ensure a representative experience:}
Overall, participants were able to correctly identify sound effects in the soundscapes; however, there were repetitive mentions about these being a bit dissonant with the artwork as they lack accuracy in terms of the period and place portrayed in the pieces. For instance, B4 mentioned, when describing their experience with the piece `Paris a Rainy Day' (painting 7), ``I love the swish of the umbrella going up and down. Probably not historically accurate. Like, they probably wouldn't have had those particular types of metal nylon umbrellas back then''. Therefore, in order to ensure a representative experience soundscapes should:
\begin{itemize}
    \item employ sound effects that are historically and contextually accurate with respect to the artwork, and
    \item use music that reflects the composition styles of the period and place were the artwork is taking place.
\end{itemize}
\subsubsection{Include ambient sound effects to aim for a more immersive experience:}

In our study experiences containing information on both the foreground and setting make it more likely that the user felt immersed in the artwork and thus had a better aesthetic experience. Hence, upcoming systems should strive to incorporate ambient elements like the gentle breeze in an open field or the rustling of trees in a forest to provide users with a more immersive experience within the artwork.

\subsubsection{Employ methods of audio reproduction for sound effects in order to enhance objective interpretation:} Sound effects should be used to convey more than just the object/scene they depict. For example, stereo information should be used to place objects within the painting or describe movement across the painting as stated by L5 "well stereo gives you direction and volume. If you're watching a game of tennis...you're gonna hear it [the ball] travel across the field". Additionally, effects such as reverb can be used to place a sound effect in the correct setting. For example, one would expect a ball bouncing in an empty, large hall to have strong reverberation. In our study, these little details had significant effects on participants' experiences and therefore should not be disregarded in future systems
 
\section{Conclusion}

In this paper, we investigated how automated musical soundscapes could be used to improve the accessibility of visual art for BLV people. Our results demonstrate that these musical soundscapes effectively create an aesthetic experience for the user, providing motivation for further research in this area. Our study also outlines a set of design considerations that can be used to guide the development of future systems and that indicate potential avenues of work for CC researchers at the intersection of fields such as narrative generation \cite{gervas2018targeted,concepcion2019evolving} and accessibility.

\bibliographystyle{iccc}
\bibliography{main}

\end{document}